\documentclass[]{spie}  

\usepackage{amsmath,amsfonts,amssymb}
\usepackage{graphicx}
\usepackage[colorlinks=true, allcolors=blue]{hyperref}
\usepackage{float}
\usepackage{tabularx}
\usepackage{booktabs}
\usepackage[labelfont={color=black,bf}]{caption}
\usepackage{subcaption}
\usepackage[dvipsnames]{xcolor}
\usepackage{hyperref}
\usepackage{orcidlink}

\title{Dealing with Segmentation Errors in Needle Reconstruction for MRI-Guided Brachytherapy}

\author[a]{Vangelis Kostoulas \orcidlink{0000-0001-5011-8195}}
\author[b]{Arthur Guijt \orcidlink{0000-0002-0480-2129}}
\author[a]{Ellen M. Kerkhof \orcidlink{0000-0002-6070-7732}}
\author[c,d]{Bradley R. Pieters \orcidlink{0000-0002-7427-8836}}
\author[b]{\linebreak Peter A.N. Bosman \orcidlink{0000-0002-4186-6666}}
\author[a]{Tanja Alderliesten \orcidlink{0000-0003-4261-7511}}
\affil[a]{Dept. of Radiation Oncology, Leiden University Medical Center, The Netherlands}
\affil[b]{Evolutionary Intelligence Group, Centrum Wiskunde \& Informatica, The Netherlands}
\affil[c]{Dept. of Radiation Oncology, Amsterdam University Medical Centers, University of Amsterdam, The Netherlands}
\affil[d]{Cancer Center Amsterdam, Imaging and Biomarkers, the Netherlands}

\authorinfo{Further author information: (Send correspondence to Tanja Alderliesten or Vangelis Kostoulas)\\Tanja Alderliesten: E-mail: T.Alderliesten@lumc.nl,\\ Vangelis Kostoulas: E-mail: E.Kostoulas@lumc.nl}

\graphicspath{ {./images/} }

\begin{document} 
\maketitle

\begin{abstract}
Brachytherapy involves bringing a radioactive source near tumor tissue using implanted needles. Image-guided brachytherapy planning requires amongst others, the reconstruction of the needles. Manually annotating these needles on patient images can be a challenging and time-consuming task for medical professionals. For automatic needle reconstruction, a two-stage pipeline is commonly adopted, comprising a segmentation stage followed by a post-processing stage. While deep learning models are effective for segmentation, their results often contain errors. No currently existing post-processing technique is robust to all possible segmentation errors. We therefore propose adaptations to existing post-processing techniques mainly aimed at dealing with segmentation errors and thereby improving the reconstruction accuracy. Experiments on a prostate cancer dataset, based on MRI scans annotated by medical professionals, demonstrate that our proposed adaptations can help to effectively manage segmentation errors, with the best adapted post-processing technique achieving median needle-tip and needle-bottom point localization errors of $1.07$ (IQR $\pm 1.04$) mm and $0.43$ (IQR $\pm 0.46$) mm, respectively, and median shaft error of $0.75$ (IQR $\pm 0.69$) mm with 0 false positive and 0 false negative needles on a test set of 261 needles.
\end{abstract}


\section{Introduction}
In brachytherapy, cancer is treated by temporarily placing a radioactive source near the tumor using dose delivery devices. For prostate cancer, interstitial needles are used. After needle implantation, a 3D image is taken, and needles are manually reconstructed by medical professionals using specialized software. This process can be challenging depending on the image quality and the patient anatomy, and time-consuming, especially as the number of needles increases. Minimizing the interval between image acquisition and dose delivery is crucial to save workload, to minimize anatomical changes, and to improve patient comfort.

\begin{figure}[H]
\centering
\includegraphics[width=\textwidth]{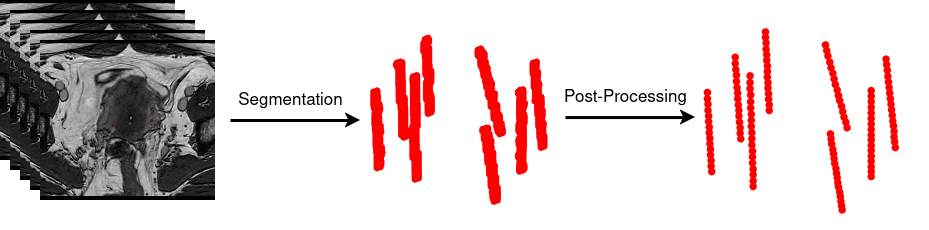}
\caption{\footnotesize{The two-stage pipeline in needle reconstruction. Images are fed into a segmentation model which is trained to predict needle point clouds. A post-processing technique is then applied to the point clouds to extract needle locations.}}
\label{fig:fig1}
\end{figure}

Over the past decade, automatic needle reconstruction pipelines have been proposed, typically comprised of two stages: 1) needle segmentation, followed by 2) post-processing to extract needle points \cite{jung2019deep, weishaupt2022approaching, kitner2023multi, liu2022challenges, shaaer2022deep, zhang2020automatic, mandal2024using} (\autoref{fig:fig1}). Deep learning models are very useful for the first stage, but their results will often contain errors, such as small false-positively identified needle parts, under-segmentation (i.e., missing a whole needle, or a part of a needle), over-segmentation (i.e., predicting a non-existent needle, or falsely extending one), and disconnected parts of needles. The second stage often fails to account for all of these errors. This is however essential in order to reduce the manual modifications that are required to be done by medical professionals. Additionally, only a few studies focused on MRI scans \cite{dai2020automatic, shaaer2022deep}, which are challenging because needles are less visible on this image modality, compared to CT scans or ultrasound images used in other works.

In \citenum{jung2019deep}, post-processing involves two steps: initialization and optimization by expectation maximization. Needles are identified by clustering deep learning predictions in 2D, followed by fitting 3D curves to these clusters, and fine-tuning the curves with expectation maximization. Segmentation errors like small false-positively identified needle parts pose problems for this post-processing technique. Moreover, needle bottom locations are assumed to be manually specified and to originate from a specific slice. This is because this post-processing technique was designed specifically for prostate cancer brachytherapy, where all the needles are implanted through a template. This technique is therefore not applicable to all needle reconstruction tasks. 

In \citenum{weishaupt2022approaching}, a post-processing technique utilizing 3D clustering with HDBSCAN is proposed. Wrongly merged clusters are detected and corrected, and polylines are fitted to the resulting clusters. Tuning HDBSCAN can be challenging, e.g., setting the minimum number of samples per cluster too low can result in too many needles, while setting it too high can merge separate needles. Moreover, this post-processing technique is also not designed to handle all potential segmentation errors (e.g., dealing with false-positively identified needle parts, disconnected needle parts, etc.).

In this paper, we aim to overcome the issues outlined above by proposing adaptations to the existing post-processing techniques identified above. In doing so, we aim to make in particular the post-processing-technique in \citenum{jung2019deep} more general, potentially extending its use beyond prostate cancer. We furthermore generally aim to introduce steps to handle segmentation-based errors to improve the overall reconstruction procedure.

\section{Methods}

\subsection{Datasets} 
We conducted retrospective experiments on two clinical datasets from the Amsterdam University Medical Center, containing 20 and 14 turbo spin echo MRI scans (Ingenia 3T Philips Healthcare, Best, The Netherlands) of prostate cancer patients, respectively, with corresponding needle annotations. The in-plane resolution for the first dataset was 0.6 $\times$ 0.7 mm. The slice thickness was 3.0 mm with a 0.3 mm gap. For the second dataset it was 0.52 $\times$ 0.52 mm with a slice thickness of 3.0 mm and a 0.3 mm gap. All the images were resampled to achieve a voxel spacing of 1.0 mm in the axial dimension. The first dataset contains 293 Ella-CS 6F needles (Ella-CS, Hradec Králové, Czech Republic), while the second dataset contains 261 Nucletron ProGuide Sharp 6F needles (Nucletron, Veenendaal, The Netherlands). 

\subsection{Label Creation} 
\begin{figure}[H]
\centering
\includegraphics[width=\textwidth]{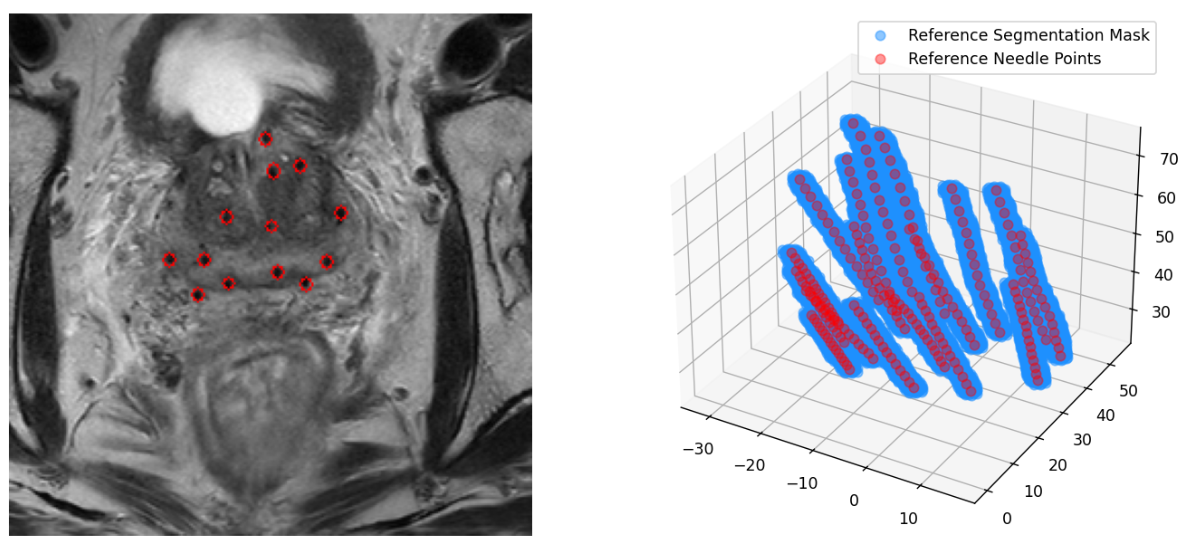}
\caption{\footnotesize{Example of the dataset. The left panel displays an axial slice with needle contours, while the right panel shows a 3D visualization of needle point clouds. The red points (reference needle points) represent needle points resulting from annotation in clinical practice, after removing bottom slices, and sampling along the needle trajectories as part of our label creation method. The blue points (reference segmentation mask) are generated by applying a dilation operation to the red points.}}
\label{fig:fig2}
\end{figure}

Some of the axial MRI slices were not manually annotated even though needles were present, as the corresponding needle locations were not needed for radiation dose delivery. We removed axial bottom slices where needles are visible but not annotated. For the remaining needle locations, we apply an interpolation on the linear segments defined by each needle's manually annotated points, followed by a dilation with a spherical kernel of radius 1 mm to increase the needle width and create reference segmentations that are better resembling their actual appearance in the images. An example of the resulting labels is illustrated in \autoref{fig:fig2}.

\subsection{nnU-Net for Needle Segmentation}
Choosing a neural network architecture and hyperparameters is challenging and time-consuming, with the best choice often being specific to the application at hand. The approach known as nnU-Net \cite{isensee2021nnu} was introduced to overcome this issue for image processing tasks. nnU-Net employs a set of heuristics to determine all hyperparameters based on the specifics of a dataset. It has been shown that state-of-the-art performance can be achieved with the use of nnU-Net on a variety of tasks, even outperforming manually constructed task-specific solutions. For this reason, we opted to use nnU-Net in this work.

Most works on needle reconstruction use 2D or 2.5D segmentation models \cite{jung2019deep, weishaupt2022approaching, liu2022challenges, shaaer2022deep, zhang2020automatic, mandal2024using}, claiming that small datasets (around 20 patients) are insufficient for training 3D models. Both 2D and 3D models are available in the nnU-Net code, so we performed training with both options to compare performance of using 2D and 3D models.

The training set is split into 5 folds in nnU-Net and 5-fold cross-validation is performed. The models are trained for 1000 epochs, and the final versions are used to form an ensemble that constitutes the final model.

The first dataset was used to train nnU-Net (16 training and 4 validation patients for each run of the 5-fold training), while the second dataset of 14 MRI scans served as the test set.

\subsection{Post-Processing}
\label{postprocessingsection}
This section outlines the post-processing techniques that we implemented in this work, both the existing techniques and our proposed adaptations. In Section~\ref{JungSection} we discuss the post-processing technique introduced by \citenum{jung2019deep}, while in Section~\ref{LeonSection} we discuss the post-processing technique proposed by \citenum{weishaupt2022approaching}. In Section~\ref{MJungSection}, we present a version of the post-processing technique of \citenum{jung2019deep} with a modified initialization step. Finally, we present adaptations to handle segmentation errors, applied to the modified post-processing technique of \citenum{jung2019deep} and the existing post-processing technique of \citenum{weishaupt2022approaching} in Sections~\ref{MJungPlusSection} and~\ref{LeonPlusSection}, respectively.

\subsubsection{Jung}
\label{JungSection}
We implemented the post-processing technique proposed by \citenum{jung2019deep}, which, as mentioned in the introduction, consists of two steps: initialization and optimization.

In the initialization step, 2D spectral clustering is employed to group predicted needle points on the most inferior axial slice. The number of clusters is predefined based on the known number of needles. For each subsequent slice, the cluster centers from the previous slice are used to group the needle points in the current slice. This process continues iteratively until one of two conditions is met: either at least two clusters converge to a very small distance (to prevent cluster merging, we used 1.5 voxels), or the last slice is processed.

Once 3D needle clusters are obtained, a polynomial is fitted to each cluster using the least squares method to initialize the needles for the subsequent optimization step. In this step, an objective function is defined as the average distance between all predicted needle points and their nearest needle. Expectation maximization is then used to minimize this function by alternating between two operations: redefining clusters by assigning points to the closest polynomial, and refitting polynomials. This process repeats until the loss function stops decreasing, or until the maximum number of iterations is reached. We used a maximum of 100 iterations.

\subsubsection{Leon}
\label{LeonSection}
As mentioned in the introduction, the post-processing technique proposed by \citenum{weishaupt2022approaching} begins with 3D clustering using HDBSCAN. In \citenum{weishaupt2022approaching}, a minimum number of samples of 15 was used and a minimum cluster size of 500. These hyperparameters depend on the specifics of the problem, such as the width of the segmented needles. In our work, we used a minimum number of samples of 5 and a minimum cluster size of 15, which we empirically determined to achieve plausible results most of the times.

To address cases where multiple needles may have been erroneously merged within a cluster, the algorithm evaluates the spatial distribution of voxels on each slice. If an abnormally large spread is detected in a slice, it is temporarily excluded from the cluster. When such regions are small, linear fitting is used to interpolate the missing segment, preserving the continuity of a single needle. For larger excluded regions, the algorithm assumes the presence of two distinct needles and applies a linear fitting technique to reconstruct trajectories for the separate fragments. This procedure results in the final clusters, each of them representing a distinct needle.

Finally, a polyline is fitted to each cluster to represent the needles. While the exact polyline fitting method used in \citenum{weishaupt2022approaching} is not specified, we utilized linear tree regression in our implementation, which we found to be effective.

\subsubsection{MJung}
\label{MJungSection}
We propose modifications to the initialization step of the post-processing technique in \citenum{jung2019deep} to eliminate the assumption of a specific starting slice for needles.

We first detect the lowest and highest points of the predicted needle point clouds in the axial dimension by applying convolutions with different kernel sizes ((3, 5, 5) and (3, 7, 7)) on the needle mask. Then, we cluster these points based on a neighborhood of radius 3.5 voxels and extract their median point. The extracted end-points are then used as references for a subsequent clustering process, performed along both axial directions (inferior-to-superior and superior-to-inferior). Starting from the most inferior or superior slice, needle points of the predicted point clouds are clustered into potential needle clouds based on the 3D distance to their closest end-point. The centroids of the resulting clusters are added to a collection, and subsequently used as the references for the clustering of the consecutive slice. Similarly to \citenum{jung2019deep}, this process continues until at least two clusters achieve a very small distance (to prevent cluster merging) or until the last slice has been reached. The initialization direction (inferior to superior or superior to inferior) which results in the largest collection of points is chosen. We refer to the post-processing technique of \citenum{jung2019deep}, but with the adapted process described above for initialization, as MJung (Modified Jung's post-processing technique).

\subsubsection{MJung+}
\label{MJungPlusSection}
While MJung is more general in that there is no longer an assumption on a specific starting slice for needles, this post-processing technique is still not robust to segmentation errors, as it automatically identifies the number of needle clusters, which could contain false positives. To overcome such limitations, we propose further adaptations.

\paragraph{Needle Merging} After initialization, we search for possible disconnected parts of needles to merge. A potential pair is identified if the bottom point of one needle lies higher in the axial direction than the top point of another needle within a 30 voxel neighborhood in the sagittal and coronal directions. For each pair, we run the RANSAC algorithm \cite{fischler1981random}  to check if a 2D polynomial can be fitted without points deviating more than 5 voxels from the fitted curve (outliers). Multiple needles can also be merged if all their pairwise combinations are potential pairs and the RANSAC algorithm finds no outliers for any combination.

\paragraph{Iterative Removal of Needles} If the number of detected needles exceeds the actual number of needles that were used (which is known), an iterative removal process is applied. This process removes one needle at a time based on the increase in total loss, defined as the average distance of all predicted needle points to their nearest needle. In each iteration, the needle that results in the smallest increase in loss is removed. This continues until the number of detected needles matches the expected count.

\begin{figure}[H]
\begin{subfigure}{.2\textwidth}
  \centering
  \includegraphics[width=.99\linewidth]{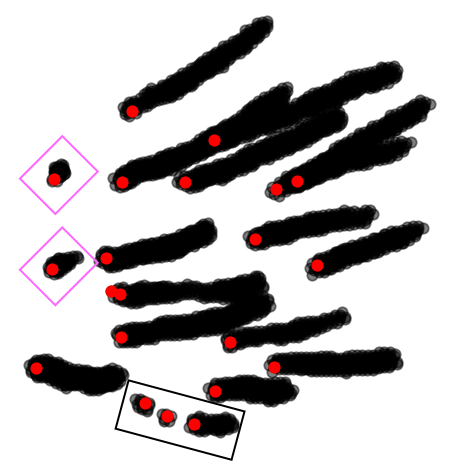}
  \caption{}
  \label{fig:sfig11}
\end{subfigure}%
\begin{subfigure}{.2\textwidth}
  \centering
  \includegraphics[width=.99\linewidth]{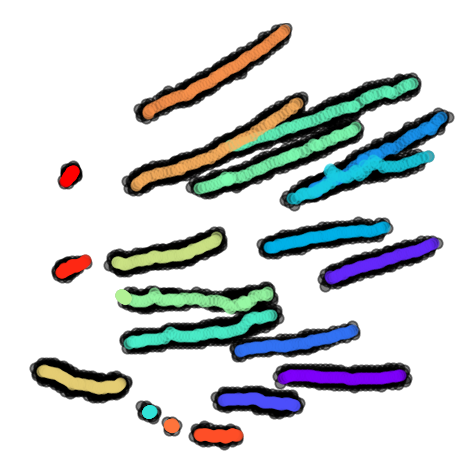}
  \caption{}
  \label{fig:sfig12}
\end{subfigure}
\begin{subfigure}{.2\textwidth}
  \centering
  \includegraphics[width=.99\linewidth]{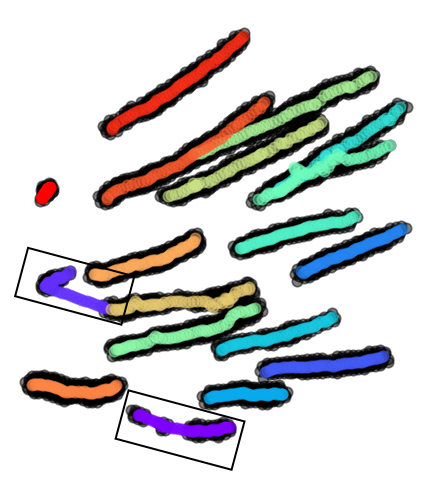}
  \caption{}
  \label{fig:sfig13}
\end{subfigure}%
\begin{subfigure}{.2\textwidth}
  \centering
  \includegraphics[width=.99\linewidth]{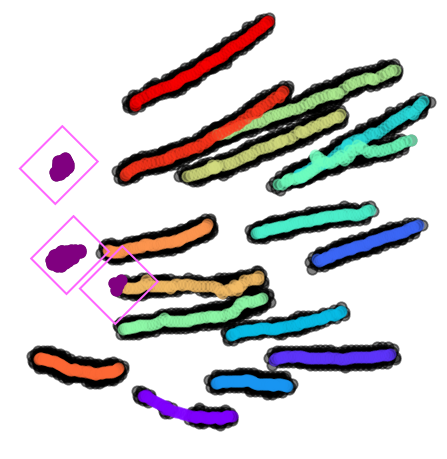}
  \caption{}
  \label{fig:sfig14}
\end{subfigure}%
\begin{subfigure}{.2\textwidth}
  \centering
  \includegraphics[width=.99\linewidth]{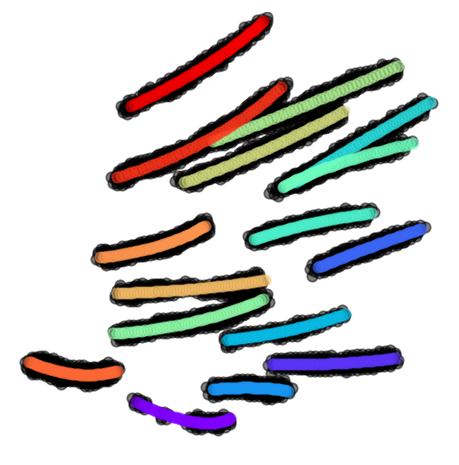}
  \caption{}
  \label{fig:sfig15}
\end{subfigure}%
\vspace*{2 mm}
\caption{Example of the main steps of MJung+. The coloring of the needles is not representative of anything but the identification of one point cloud (one needle). The points predicted by the deep learning model are colored black. In step (a) the bottom points of needles are extracted (red points). With the black box we indicate a case of a needle split in three parts, and the purple boxes highlight two false-positively identified needle parts. In (b) the result of running the modified initialization of \citenum{jung2019deep} is illustrated. In (c) the result of merging needles is illustrated. Notice that the merging algorithm merged the three disconnected parts of a needle, but it also merged two false-positively identified needle parts (black boxes), highlighted in the figure of the next step. In (d) the result is shown of the detection and subsequent removal of false-positively identified needle parts, through the iterative needle removal process described above, highlighted with purple boxes. Finally, in (e) the result of running the optimization step to get the final needles is shown.}
\label{fig:fig3}
\end{figure}

\paragraph{Polynomial Fitting and Optimization}
Finally, for each resulting needle, we fit a polynomial using least squares to obtain the initial needles for a subsequent optimization step (expectation maximization). We adopt the same objective function as \citenum{jung2019deep} and perform expectation maximization until the loss stops decreasing or the maximum number of iterations is used. Similar to Jung and MJung, we used a maximum of 100 iterations. Unlike \citenum{jung2019deep}, we run the optimization step three times with different polynomial degrees (1, 2, and 3) and select the polynomial fit for which the smallest loss was found. \autoref{fig:fig3} illustrates all the steps of MJung+.

\subsubsection{Leon+}
\label{LeonPlusSection}
After running the post-processing technique described in Section~\ref{LeonSection}, we can subsequently also use the adapted and additional steps as we proposed in Section~\ref{MJungPlusSection}.

\subsection{Evaluation}
We compared all discussed post-processing techniques above, tuning all their hyperparameters on the validation sets of the 5-fold training. The results of this tuning are the settings we have described above in Section~\ref{postprocessingsection}.

For each post-processing technique, we sampled 100 equidistant points along the trajectory of each needle found by the technique, to calculate metrics. We calculated the median tip and bottom point localization errors as well as the median shaft error. The latter is the average distance between corresponding pairs of points (i.e., looking at points with the same index resulting from the sampling of 100 equidistant points along the needles) of the predicted needle and the manually annotated needle. We also counted the number of needles with a shaft error greater than 1 mm and the number with a shaft error greater than 2 mm as well as the number of false positive needles that were detected.

\section{Results}
\begin{figure}[H]
\centering
\includegraphics[width=0.6\textwidth]{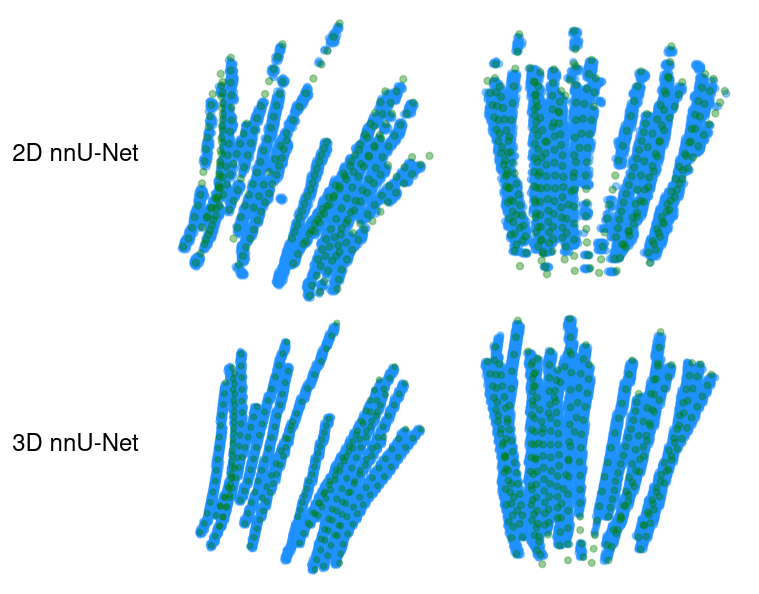}
\caption{\footnotesize{Examples of the segmentation results of nnU-Net. Points predicted using nnU-Net are colored blue, while the green points are the labels created based on the clinical needle points. The 2D nnU-Net produced many more false identified needle parts and disconnected needles than 3D nnU-Net.}}
\label{fig:fig4}
\end{figure}

In \autoref{table:table1}, the results of our experiments are presented. Despite being trained with different needles, nnU-Net successfully detected the needles in the test set. However, we observed a very large difference between the 2D and 3D segmentation models. The 2D model produced significantly more false positively identified needle parts, and discontinuous needles (see \autoref{fig:fig4}). 

\begin{table}[H]
\centering
\caption{Median localization errors and IQR in mm for needle tips, bottom points, and shaft error on the test set (261 needles) for the 5 different post-processing techniques and for the 2D and 3D nnU-Net models. The number of needles found (NF), needles with shaft error larger than 2 mm (NSEB2) and 1 mm (NSEB1), and  false positives (FP) are also reported. $\downarrow$ and $\uparrow$ mean lower or higher values are better, respectively.}

\tabcolsep=1.7mm
\ \\\ \\\begin{tabular}{|c|c|c|c|c|c|c|c|c|} 
  nnU-Net & Technique & Tip Error $\downarrow$ & Bottom Error $\downarrow$ & Shaft Error $\downarrow$ & NF $\uparrow$ & NSEB2 $\downarrow$ & NSEB1 $\downarrow$ & FP $\downarrow$ \\ [0.5ex] 
 \hline
  & Jung \cite{jung2019deep} & $1.11 \pm 1.14$ & $0.42 \pm 0.44$ & $0.75 \pm 0.71$ & 255 & 48 & 85 & 0 \\ 
  & Leon \cite{weishaupt2022approaching} & $1.02 \pm 1.07$ & $0.37 \pm 0.34$ & $0.72 \pm 0.66$ & 261 & 48 & 83 & 6 \\
 3D & MJung & $1.05 \pm 1.04$ & $0.42 \pm 0.44$ & $0.74 \pm 0.70$ & 261 & 47 & 86 & 8 \\
  & MJung+ & $1.07 \pm 1.04$ & $0.43 \pm 0.46$ & $0.75 \pm 0.69$ & 261 & 45 & 85 & 0 \\
  & Leon+ & $1.05 \pm 1.03$ & $0.43 \pm 0.45$ & $0.75 \pm 0.69$ & 261 & 45 & 85 & 0 \\  
 \hline
  & Jung \cite{jung2019deep} & $1.80 \pm 2.68$  & $0.49 \pm 0.50$  & $1.10 \pm 1.68$  & 193 & 59  & 107 & 0 \\ 
  & Leon \cite{weishaupt2022approaching} & $4.46 \pm 25.52$ & $0.92 \pm 33.30$ & $6.14 \pm 18.18$ & 261 & 160 & 192 & 372 \\
 2D & MJung          & $2.88 \pm 13.80$ & $0.90 \pm 12.96$  & $2.85 \pm 13.50$ & 236 & 139 & 167 & 196 \\
  & MJung+           & $1.99 \pm 4.52$  & $0.74 \pm 5.18$  & $1.72 \pm 5.41$  & 235 & 111 & 156 & 0 \\
  & Leon+           & $1.92 \pm 5.00$  & $0.75 \pm 10.62$ & $1.96 \pm 8.99$ & 261 & 129 & 173 & 0 \\
 
\end{tabular}
\label{table:table1}
\end{table}

The Jung post-processing technique missed a considerable number of needles and was heavily impacted by segmentation errors. Similarly, the Leon post-processing technique could not handle segmentation errors, detecting 372 false positives on the 2D nnU-Net results. This large number of false positives was partly due to the tuning of HDBSCAN, but increasing the minimum cluster size and number of samples caused this technique to merge or miss multiple needles, and thus achieve worse performance.

\begin{figure}[H]
\centering
\includegraphics[width=0.9\textwidth]{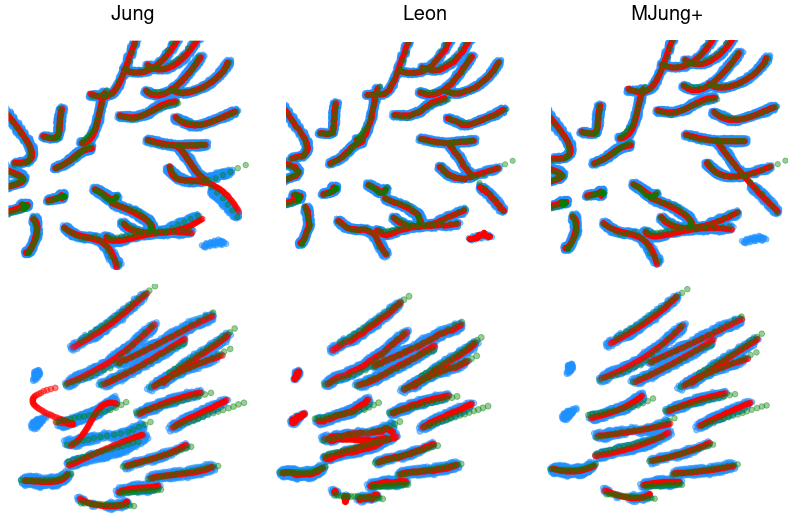}
\caption{\footnotesize{Examples of results from using 3D nnU-Net and a selection of the post-processing techniques. The points predicted from the deep learning model are colored blue, the final needles from the post-processing techniques are red, and the green points represent the manually defined points by medical professionals. The first example (top row) shows that Jung's technique is not able to connect two disconnected needle parts, and one needle is impacted by a false identified needle part. Leon's technique cannot connect the disconnected needle parts and consequently falsely identifies a needle part as a needle. Similarly, in the second example (bottom row) Jung's technique is impacted by falsely identified needle parts. Leon's technique also is impacted but additionally it cannot connect three disconnected needle parts.}}
\label{fig:fig5}
\end{figure}

While our proposed adaptations of the Jung technique to obtain the MJung technique enable overcoming the assumption of having a common needle plane, the results indicate that they also lead to detecting more false positives, as MJung generally cannot handle false positively identified needle parts. By adding our steps to address segmentation errors (MJung+), we reduced the false positives to zero. Overall, all post-processing techniques achieve similar quantitative results, but the post-processing techniques MJung+ and Leon+ lead to fewer false positives and fewer needles with shaft errors $>$ 1 mm and $>$ 2 mm. Additionally, the latter post-processing techniques are the only ones capable of handling false positively identified needle parts and disconnected needles properly (see \autoref{fig:fig5}).

\section{Discussion}
In this work, we compared post-processing techniques for needle reconstruction based on automated needle segmentations in MRI scans of prostate cancer patients. We compared both existing techniques and newly proposed adaptations thereof.

We modified the post-processing technique in~\citenum{jung2019deep} by extending its use to cases where needles do not originate from a specific slice. Moreover, we proposed additional steps that can be used in post-processing techniques to handle segmentation errors. In this work, we showed that with our proposed adaptations, the performance of the modified post-processing technique in~\citenum{jung2019deep} and the post-processing technique in~\citenum{weishaupt2022approaching} was improved.

Potentially, the incorporation of our proposed adaptations and added steps can also benefit other, including future, post-processing techniques, as long as they employ a form of clustering to identify distinct point clouds that may represent needles.

For the optimization step, a different optimization algorithm, or curve fitting technique could be used. In our work, we found that expectation maximization for optimization as used in \citenum{jung2019deep} performed well enough, and we left experiments with other optimization algorithms for future research. In \citenum{weishaupt2022approaching}, no optimization step is included, and polyline fitting is performed. In our experiments, we found that choosing polyline fitting over polynomial fitting does not make a big difference in terms of quantitative results. However, we noticed that polynomial fitting results in smoother needle trajectories and, therefore, more realistic needles, which is why we ultimately chose to use polynomial fitting in our proposed adaptations of the modified post-processing technique in~\citenum{jung2019deep} and the post-processing technique in~\citenum{weishaupt2022approaching}.

For the automated segmentation, we used deep learning, specifically nnU-Net. We trained both 2D and 3D segmentation models to investigate their differences when used as input for the post-processing techniques. 

The results indicate that using a 3D segmentation model based on nnU-Net trained with MRI scans significantly reduces false positives and discontinuous needles compared to using its 2D counterpart, even when using a training dataset of only 16 patients. Moreover, extending post-processing techniques with our steps to address segmentation errors leads to improved results and more effective management of segmentation errors.

\section{Conclusions}
The most common approach to automate the reconstruction of needles from medical images in brachytherapy is to use an (AI-based) segmentation model, followed by a post-processing technique to derive the needle configurations from the segmentation masks. Existing approaches to the post-processing part do not effectively address segmentation errors, which are common in deep learning models. We have proposed adaptations to existing post-processing techniques that specifically target potential errors in segmentation output and have shown that our proposed adaptations can often effectively handle such segmentation errors, making the overall automated reconstruction procedure more reliable. However, to assess the clinical usefulness of these tools, a validation study in which medical professionals evaluate the outcomes of the final techniques still needs to be performed.


\section*{ACKNOWLEDGMENTS}
\label{sec:acknowledgements}

The research is part of the research programme Open Technology Programme with project number 18373, which is financed by the Dutch Research Council (NWO), Elekta, and ORTEC LogiqCare.

\bibliography{4_pages} 
\bibliographystyle{spiebib} 

\end{document}